\documentclass[aps,prl,amsmath,amssymb,reprint,superscriptaddress]{revtex4-1}
\usepackage{graphicx}
\usepackage{amsmath}
\usepackage{dcolumn}
\usepackage{bm}
\usepackage{bbold}
\usepackage{color}
\usepackage{amsfonts}
\usepackage{amssymb}
\usepackage{mathrsfs}
\usepackage{braket}

\usepackage{caption}
\usepackage{subcaption}

\begin{document} 

\title{Spin-Photon Interaction in a Cavity with Time-Reversal Symmetry Breaking}

\author{Maxim Goryachev}
\email{maxim.goryachev@uwa.edu.au}
\affiliation{ARC Centre of Excellence for Engineered Quantum Systems, School of Physics, University of Western Australia, 35 Stirling Highway, Crawley WA 6009, Australia}

\author{Warrick G. Farr}
\affiliation{ARC Centre of Excellence for Engineered Quantum Systems, School of Physics, University of Western Australia, 35 Stirling Highway, Crawley WA 6009, Australia}

\author{Daniel L. Creedon}
\affiliation{ARC Centre of Excellence for Engineered Quantum Systems, School of Physics, University of Western Australia, 35 Stirling Highway, Crawley WA 6009, Australia}

\author{Michael E. Tobar}
\affiliation{ARC Centre of Excellence for Engineered Quantum Systems, School of Physics, University of Western Australia, 35 Stirling Highway, Crawley WA 6009, Australia}

\date{\today}


\begin{abstract}

Employing a sapphire whispering gallery mode resonator, we demonstrate features of the spin-photon interaction in cavities with broken time-reflection symmetry. The broken symmetry leads to a lifting of the degeneracy between left-handed and right-handed polarised cavity photons, which results in an observable gyrotropic effect.  In the high-$Q$ cavity limit, such a situation requires a modification of the Tavis-Cummings Hamiltonian to take into account conservation of spin angular momentum and the corresponding selection rules. As a result, the system is represented by a system of two linearly coupled bosonic modes, with each one coupled to its own sub-ensemble of two-level systems with different energy splittings. In the experimental example, these sub-ensembles originate from Fe$^{3+}$ impurity ions effectively seen as a two level systems at the interaction frequency. The temperature dependence of the population of each sub-ensemble (in terms of effective susceptibility of the medium) is determined experimentally in accordance with the theoretical predictions revealing various paramagnetic impurity types in the solid. The regimes of backscatterer and spin ensemble domination are discussed and compared. 

\end{abstract}

\maketitle

\section{Introduction}

The Jaynes-Cummings Hamiltonian\cite{JC} is an important model describing the interaction of a two level quantum system with an electromagnetic cavity mode. The cavity is usually understood as a linear cavity of the Fabry-P\'{e}rot type, where the electromagnetic wave propagating as $e^{\pm ikx}$ is confined between two mirrors. Such a cavity explicitly possesses { both time-reversal (T-symmetry) and reflection symmetries}. It is typically implied that the confined light is linearly polarised\cite{SCULLY} and thus effectively has no spin angular momentum. Semi-classically this can be understood as a superposition of left and right hand propagating electromagnetic waves that form a standing wave and interact with an atom. These waves are always degenerate in frequency and are mirror reflections of each other, thus the matter of spin angular momentum conservation becomes irrelevant due to symmetry under the mirror $x\mapsto-x$ {as well as time reversal $t\mapsto-t$ transforms}. 

In a circular cavity where a wave travels around the cavity boundary $e^{\pm ik\phi}$, such as a Whispering Gallery Mode (WGM) type resonator, there is continuous rotational symmetry $\phi\mapsto\phi+\theta$. Similarly to a Fabry-P\'{e}rot cavity, the modes of such systems can be represented as linear combinations of clockwise and counter-clockwise circularly polarised waves. However, conversely to the linear cavity, the reflection symmetry of the system ($\phi\mapsto-\phi$) may be broken. For example, back-scatterers in a cavity lift the degeneracy by introducing a coupling between the two counter-propagating waves, causing the mode to appear as a doublet. {In this case of disorder symmetry breaking,} the linear combination of the two counter propagating modes causes spatially orthogonal standing waves\cite{gorod,weiss,giordano}. However, in the case where the dominant defect in the cavity is an ion with spin, a gyrotropic response { with the underlining T-symmetry breaking} has previously been observed between the two standing wave modes when the ion's electron spin resonance frequency has been tuned to the WGM frequency\cite{paperA,karim1}. It has been shown that in this case, one of the waves is a clockwise travelling wave, while the other is a counter-clockwise travelling wave, which have the distinction of being right and left circularly polarised electromagnetic waves.
 It is widely known that photons of such waves carry a spin angular momentum of $\pm1$. In this work we employ the language where a photon has spin 1 if it is in the  $\Ket{L}$ state and $-1$ if it is in the $\Ket{R}$ state. So, photons in both resonances of the doublet carry a projection of the spin angular momenta (helicity) $+\hbar$ (further denoted by its polarisation state $\Ket{L}$) or $-\hbar$ (further denoted by its polarisation state $\Ket{R}$). This situation is equivalent to a system of two coupled Harmonic Oscillators (HO) confining photons with spin projection of either $+\hbar$ or $-\hbar$. 

The comparison of the two cavity types demonstrates certain differences arising due to the T-symmetry breaking phenomenon in a system of two travelling waves.  In particular, in the case of broken T-symmetry, the atom-field interaction requires the conservation of the system's spin angular momentum that is possessed by ions and carried by travelling waves, implying one more selection rule beyond energy conservation. As a result, one has to consider both energy and spin angular momentum\cite{allen,karassiov} conservation laws to describe the interaction between matter and photons. This interaction between cavity photons and an ensemble of spins\cite{ritsch} results in a three HO model and explain the observed gyrotropic effects. 

%

\section{Physical Realization}

The physical realization of a cavity with circular geometry is the well known WGM dielectric resonator, extensively studied both in the optical and microwave domains. For extremely low-loss microwave systems such as sapphire resonators, the existence of doublets is also well known~\cite{karim1,modeplit}. They arise due to various imperfections of the crystal resonators such as the presence of backscatterers, and can be effectively controlled with external magnetic field via interaction with impurity ions\cite{paperA}.

The sapphire crystals that comprise such resonators obtain naturally occurring impurity ions within the lattice during the growth procedure, such as Fe$^{2+}$, Fe$^{3+}$, Cr$^{3+}$, and V$^{2+}$\cite{sapphSPEC}. One of the most most abundant impurity ions is Fe$^{2+}$, which may be converted to Fe$^{3+}$ by annealing in an oxygen environment\cite{creedon1, creedon2}. The effects of Fe$^{3+}$ in sapphire have been studied previously both theoretically and experimentally~\cite{PhysRevB.88.235104,karim2,KorPro} and in this work we study a sample with impurity concentration of the order of 100 parts per billion. Substitutional Fe$^{3+}$ ion impurities have a $^6$S ground state with a 3d$^5$ electronic configuration of the unfilled shell. Zero-field splittings (three energy levels) of this ground state have been discussed theoretically and used experimentally to create a solid-state maser at microwave frequencies\cite{pyb1}. The Zeeman effect splits these three levels into two sub levels each in the presence of a DC magnetic field, giving rise to six quantum states of the ion: $\Ket{\pm1/2}$, $\Ket{\pm3/2}$, $\Ket{\pm5/2}$. Each of these states is described by a value of the electron spin angular momentum. Moreover, the energy level structure of Fe$^{3+}$ ions in sapphire is such that there exists ion-exciting transitions, which can either increase or decrease the spin angular momentum.  

In the present work, a cylindrical sapphire monocrystal designed to sustain WG modes in the range $5-20$ GHz was cooled in a dilution refrigerator. An external DC magnetic field was applied along the $z$-axis of the crystal in order to split the energy levels of the Fe$^{3+}$ ions in the crystal by such an amount that the frequency of two allowed transitions $\Ket{\pm1/2}\rightarrow\Ket{\pm3/2}$ corresponded to the frequencies of particular WGMs of the crystal. Each WGM used in the experiment was split into a doublet, where each resonance of the doublet confines photons of certain polarisation: right hand circular (RHC) or left hand circular (LHC) when interacting with the Fe$^{3+}$ ions. Each doublet resonance is clearly resolvable due to the extremely low photon loss ($Q> 10^8$) and correspondingly small mode bandwidths - significantly smaller than the doublet mode splitting. Because of their outstanding quality factors, WGMs in sapphire crystals with highly-dilute impurities are an excellent platform to study matter-wave interactions in the case of broken T-symmetry.

\section{System Description}

As described in the Introduction, the main consequence of breaking the symmetry during the ion-photon interaction is the existence of two non-degenerate HOs with opposite values of spin angular momentum projection. As a result, these HOs interact with matter in different ways due to an extra requirement on spin angular momentum conservation.
Thus, photons carrying a certain spin projection can only be coupled to ion transitions having a certain change of spin angular momentum. In the case of a $\Ket{L}$ photon that increases ion spin angular momentum by $\hbar$, the interaction is only possible with an energy level transition of the ion that corresponds to positive change of its spin angular momentum. 
Accordingly, photons from the other HO of the doublet do not interact with this particular transition. In the case of atoms whose higher energy state has spin angular momentum lower than that of the ground state, the opposite situation is observed. 

For Fe$^{3+}$ impurity ions in sapphire, transitions with both an increase ($\Ket{+1/2}\rightarrow\Ket{+3/2}$) and a decrease ($\Ket{-1/2}\rightarrow\Ket{-3/2}$) of spin angular momentum by $\hbar$ are possible. The former is referred to here as a spin-increasing ion transition (corresponding to the level splitting $\omega_+$ and spin change $\Delta m_S = +1$), whilst the latter as a spin-decreasing transition (corresponding to the energy level splitting $\omega_-$ and spin change $\Delta m_S = -1$). Each of these transitions within the ion corresponds to different energy splittings, and thus requires photons of different energies for interaction. As a result, one has to consider both energy and spin angular momentum selection rules to describe the interaction between matter and photons. From the Fe$^{3+}$ ion point of view, the impurity interacts with a photon of certain polarisation, depending on its initial state. For simplicity, we group all the ions into three sub-ensembles depending on the initial state: $\Ket{+1/2}$, $\Ket{-1/2}$, or other, where the number of ions in each sub-ensemble is given by their thermal distribution. As a result, the whole system can be represented by two intercoupled HOs that are separately coupled to two ensembles of Two Level Systems (TLS). Such a system can then be represented by the following generic Hamiltonian:
\begin{multline}
\label{B002SF}
\displaystyle  H = \omega_R\hbar a_R^\dagger a_R +  \omega_L\hbar a_L^\dagger a_L + g_{RL}\hbar\Big(a_R^\dagger a_L+ a_R a_L^\dagger\Big)\\
\displaystyle+\hbar\frac{\omega_+}{2}\sum_{i=1}^{N_+}\sigma^z_{+i} + \hbar\frac{\omega_-}{2}\sum_{i=1}^{N_-}\sigma^z_{-i} +\hbar \widetilde{g}_+\sum_{i=1}^{N_+}\big(\sigma_{+i}^+a_L+\sigma_{+i}^-a_L^\dagger\big)\\
\displaystyle +\hbar \widetilde{g}_-\sum_{i=1}^{N_-}\big(\sigma_{-i}^+a_R+\sigma_{-i}^-a_R^\dagger\big)
\end{multline}
where $a_R^\dagger$ ($a_R$) and $a_L^\dagger$ ($a_L$) are the creation (annihilation) operators for RHC and LHC polarised photons of a doublet, $g_{RL}$ is the coupling between RHC and LHC photon modes through the backscattering mechanism,  $\widetilde{g}_\pm=g_\pm /\sqrt{N_{\pm}}$ are the couplings between each type of photon and ion per spin, $\sigma_{\pm i}^z$, $\sigma_{\pm i}^x$ and $\sigma_{\pm i}^y$ are usual Pauli matrices approximating the behaviour of a Fe$^{3+}$ ion as an $i^{\text{th}}$ TLS (with a level splitting $\omega_\pm$) in a given sub-ensemble, i.e. in a required state:
\begin{equation}
	\label{B003SF}
	\left. \begin{array}{ll}
\displaystyle  \sigma^z_{+i} = \Ket{+3/2}\Bra{+3/2}-\Ket{+1/2}\Bra{+1/2}, \\
\displaystyle \sigma_{+ i}^+ = \Ket{+3/2}\Bra{+1/2}, \hspace{5pt} \sigma_{+ i}^- = \Ket{+1/2}\Bra{+3/2},\\
\displaystyle  \sigma^z_{-i} = \Ket{-3/2}\Bra{-3/2}-\Ket{-1/2}\Bra{-1/2}, \\
\displaystyle \sigma_{- i}^+ = \Ket{-3/2}\Bra{-1/2}, \hspace{5pt} \sigma_{- i}^- = \Ket{-1/2}\Bra{-3/2}.\\
\end{array} \right.
\end{equation}
consequently for a TLS with increasing and decreasing spin angular momentum during excitation. 
The first two terms of the Hamiltonian (Eqn. \ref{B002SF}) represent two HOs corresponding to the $\Ket{L}$ and $\Ket{R}$ photon spin states. These HOs are close in angular frequency $\omega_R\gg\big|\omega_R-\omega_L\big|$ because backscatterers exist only as small perturbations. The third term gives the linear coupling between two resonances arising, for example, due to back scatterers. The forth and fifth terms give a TLS approximation for the ensemble of impurity ions. The final term in the Hamiltonian describes the corresponding interactions between each sub-ensemble and a particular HO. In the experiment discussed here, the incident power is always kept low enough to keep the spin sub-ensembles unsaturated, so as to avoid all the nonlinear effects previously observed in sapphire WGM resonators\cite{PhysRevB.88.235104}.

\begin{figure}[ht!]
	\centering
			\includegraphics[width=0.35\textwidth]{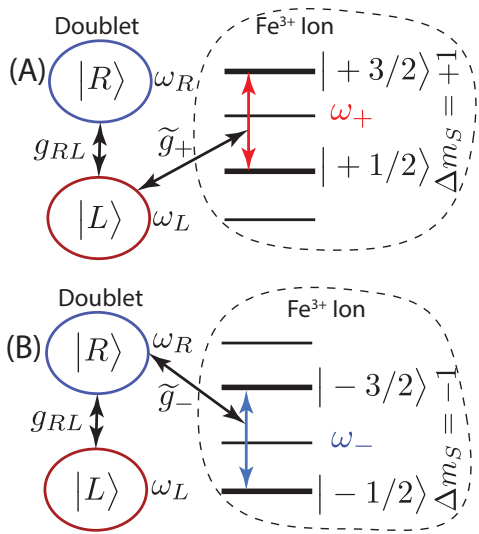}
	\caption{Energy diagram of the Fe$^{3+}$  ion interaction with a double resonance in a cavity with the broken T-symmetry. (A) shows the case of a spin-increasing ion transition where $\omega_+$ is tuned to the doublet, and (B) is the case when $\omega_-$ is tuned to the doublet.}
	\label{geom}
\end{figure}

The Hamiltonian represents the interaction between a bosonic WGM of the circular cavity with an ensemble of two level systems, and is is a modification of the well known Tavis-Cummings Hamiltonian\cite{Tavis:1968kq,ritsch,Verdu:2009nr} for the case of a cavity with broken symmetry. The Hamiltonian for the present system employs two Harmonic oscillators for the left and right circularly polarised resonances of the doublet. The two harmonic oscillators are linearly coupled to each other through the backscattering, and each resonance is coupled to the bath of TLS due to the Fe$^{3+}$ impurity ions. However, due to the spin angular momentum conservation law, only one HO of the doublet is able to interact with the impurity ions at a given transition frequency when tuned by the external magnetic field. The two possible situations are described in Fig.~\ref{geom}.  When the spin-increasing transition $\omega_+$ is tuned close to the frequency of the doublet $\omega=\frac{1}{2}(\omega_R+\omega_L)$, only $\Ket{L}$ photons are allowed to interact demonstrating a frequency shift (situation (A) in Fig.~\ref{geom}). In situation (B) of Fig.~\ref{geom}), when $\omega_-$ is tuned close to the frequency of the doublet $\omega$ only $\Ket{R}$ photons are allowed to interact demonstrating a frequency shift. In both cases, due to direct coupling between HOs, a HO that is not coupled to ions also exhibits a very small relative frequency deviation. Thus, if the HO inter-coupling is less than the spin-photon coupling ($g_{\pm}\gg g_{RL}$) and the cavity resonance linewidth, a gyrotropic effect is observed. In this case, one of the resonances is much more highly tunable by the magnetic field than the other. In contrast, when $g_{\pm}<g_{RL}$, both resonances respond in a similar manner and no gyrotropic effect is observed, a fact that was confirmed in our modelling. 

\section{Experimental Observations}

\begin{figure}[ht!]
	\centering
			\includegraphics[width=0.43\textwidth]{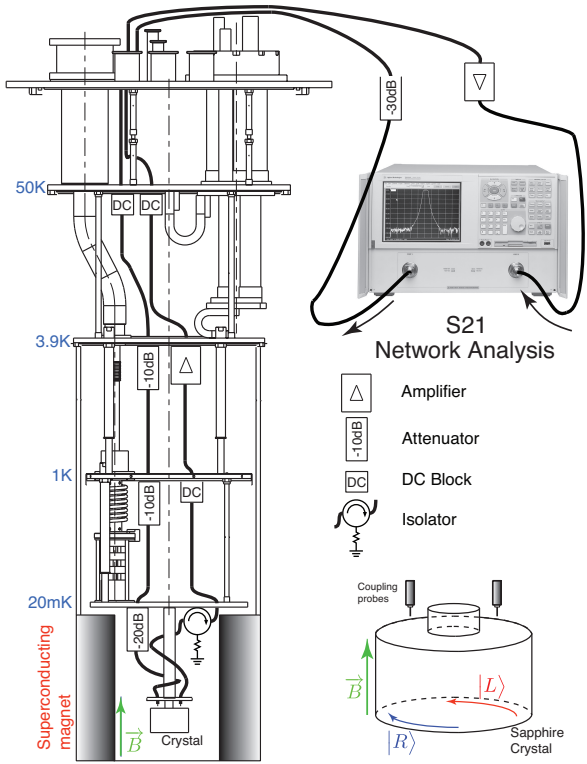}
	\caption{Experimental setup: the incident signal is attenuated at various stages of the dilution refrigerator, the transmitted signal is amplified both at 4K and at rom temperature. The backaction noise of the cold amplifier is damped by a milli-Kelvin isolator. Anti-radiation shields are not shown. }
	\label{setup}
\end{figure}

To experimentally observe the aforementioned effects, two fundamental WGMs at $f_A=13.259$~GHz (WGH$_{19,0,0}$) and $f_B=10.8104$~GHz (WGH$_{15,0,0}$) {of a cylindrical sapphire resonator were studied. The experimental setup is demonstrated in Fig.~\ref{setup}. The cylinder ($50$~mm diameter, $30$~mm height) is oriented in such a way that the $c$-axis of the crystal is parallel to the external DC magnetic field and excited with the straight antennae. The cavity transmission is characterised by measuring the scattering transmission parameters S21 with a network analyser with a cascade of cold and room temperature attenuation and amplification. }

\begin{figure}[ht!]
	\centering
			\includegraphics[width=0.50\textwidth]{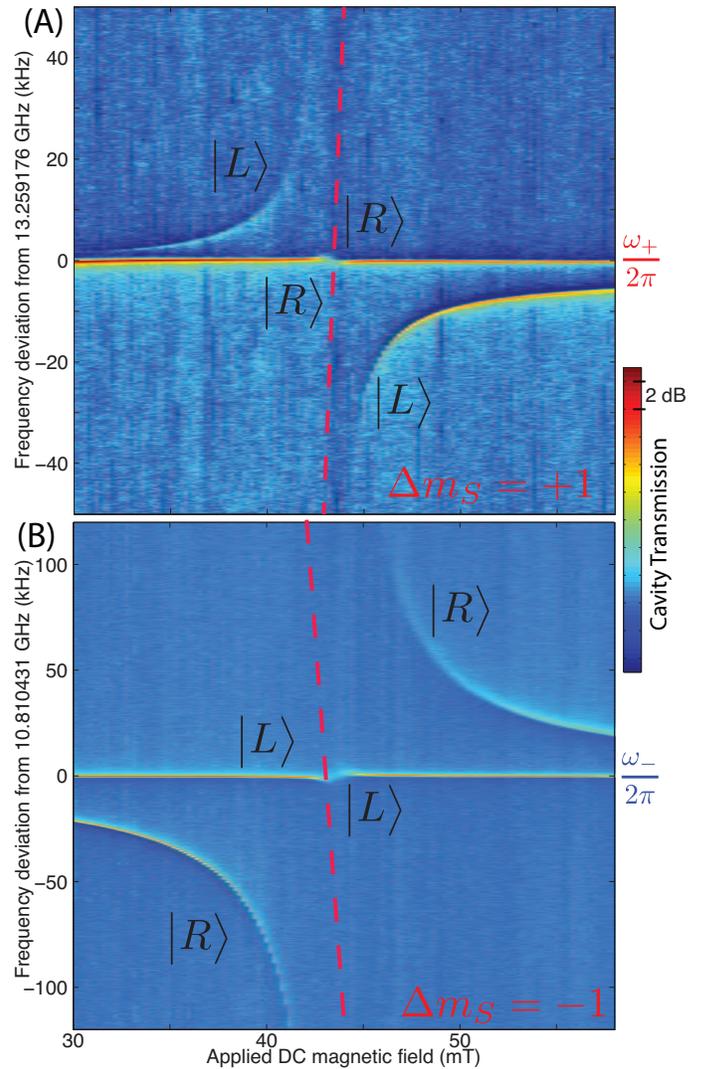}
	\caption{Interaction of the two sub-ensembles of Fe$^{3+}$ ions with two different WGM doublets at 36 mK and  $B\approx44$ mT. (A) Spin-increasing ion transition $\omega_+$ tuned to the doublet $2\pi f_A$ (13.259 GHz), where $g_+ \gg g_{RL}$ and $g_- \sim 0$.  (B) Spin-decreasing ion transition $\omega_-$ tuned to the doublet $2\pi f_B$ (10.810 GHz), where $g_- \gg g_{RL}$ and $g_+ \sim 0$.}
	\label{plus}
\end{figure}

{Interactions of $f_A$ and $f_B$} doublets with spin-increasing and decreasing transitions of Fe$^{3+}$ are shown in Fig.~\ref{plus}. In both cases only one of the HOs ($\Ket{L}$ at $f_A=\frac{\omega_+}{2\pi}$ and $\Ket{R}$ at $f_B=\frac{\omega_-}{2\pi}$) is coupled to the spin transition in accordance with the spin number conservation law. These results are obtained at low temperatures such that spin-photon coupling is much greater than coupling between HOs. In such a case, the influence of one of the polarizations is negligible, so that the whole interaction picture can be modelled using a reduced two oscillator model - one for the photonic mode and one for the spin ensemble (Fig.~\ref{plus}). The dashed lines denote a one-photon transition Zeeman effect $\omega_\pm = \frac{g\beta_e}{\hbar}B$ where $g$ is the electron $g$-factor, $\beta_e$ is the Bohr magneton and $B$ is the external magnetic field. The picture is mirror symmetric around the $B=0$ line. 

Due to the relatively large bandwidth of the Fe$^{3+}$ transitions (on the order of 25 MHz), these measurements are not in the strong coupling regime between cavity photons and spins (although the linewidth of the WG mode resonance is narrow enough). Thus, a double peak structure has been never observed for these impurities in sapphire. The best coupling between spins and photons is observed to be about 6 MHz, thus we estimate that the concentration of spins needs to be increased from about 100 parts per billion to about 2 parts per million to achieve this, which is about the level of concentration of Fe$^{2+}$ ions in the sample. Thus, if either all Fe$^{2+}$ ions could be converted to Fe$^{3+}$, or the crystal could intentionally be doped to that level, strong coupling could, in principle, be attained in the future.

\begin{figure}[ht!]
	\centering
			\includegraphics[width=0.5\textwidth]{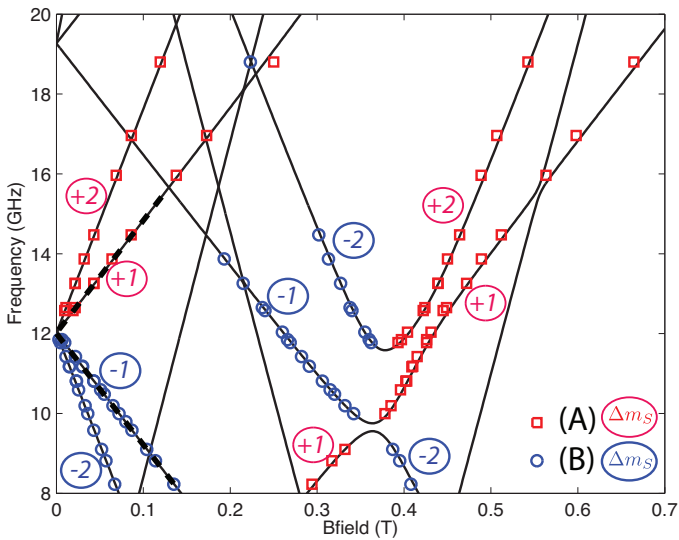}
	\caption{Interactions of the Fe$^{3+}$ ion ensemble with two WGM doublets. Solid curves are theoretical predictions. Squares denote spin-increasing transitions (situation (A) in Figs.~\ref{geom} \& \ref{plus}), and circles denote spin-decreasing transitions (situation (B) in Figs.~\ref{geom} \& \ref{plus}). Dashed lines are particular transitions discussed in the text.}
	\label{zeemanp}
\end{figure}

Fig.~\ref{plus} shows the interaction between the $\Ket{+1/2}\rightarrow\Ket{+3/2}$ and $\Ket{-1/2}\rightarrow\Ket{-3/2}$ ion transitions and two WGMs at the specific frequencies of 13.259 and 10.810 GHz. Over a broader range of fields and in the frequency range 8-20 GHz the amount of interactions with different WGMs is quite numerous\cite{sapphSPEC}. As well as the Fe$^{3+}$ ion being a six-level system, it can exhibit other transitions too. Fig.~\ref{zeemanp} presents the full spectroscopy over this range, both experimentally (squares and circles) and theoretically (solid lines), of Fe$^{3+}$ ions in sapphire. All interactions here are sorted into two groups: spin-increasing ($\Delta m_S>0$) and spin decreasing ($\Delta m_S<0$). The change in spin number $\Delta m_S$ is shown for each part of the curves. The features of these interactions are the same as discussed above in that selection rules apply due to conservation of spin angular momentum. 



As described in the previous section, the intensity of the photon-spin interaction at $\omega_+$ or $\omega_-$ depends on how many ions from the ensemble are prepared at the corresponding `TLS ground state', $\Ket{+1/2}$ or $\Ket{-1/2}$. At finite temperatures in the continuously driven regime, this number is a function of the thermal distribution of ions.
Thus, the numbers $N_+$ and $N_-$ in Eqn.~(\ref{B002SF}) denote the number of TLS available for interaction with photons of certain polarization, i.e. the number of electrons in each sub-ensemble. Consequently, they may be approximated as the expectation value of the number of ions in the $\Ket{+1/2}$ and $\Ket{-1/2}$ states. These numbers may be calculated using statistics of the system at temperature $T=\frac{1}{k_B\beta}$, by:

\begin{equation}
\label{B004aSF}
	\begin{aligned}
		\displaystyle N_+ &= \frac{N_T}{Z}\exp\Big(-\beta E_{\Ket{+1/2}}\Big),\\
 		\displaystyle N_- &= \frac{N_T}{Z}\exp\Big(-\beta E_{\Ket{-1/2}}\Big),
	\end{aligned}
\end{equation}
where $Z$ is the corresponding impurity partition function, $E_{\Ket{\pm1/2}}$ are energies of corresponding states, and $N_T$ is the total number of impurity ions participating in the interaction with the WGMs. This means that the size of the ensemble of TLS that interacts with each photon polarisation will depend on temperature. Thus, by changing the resonator temperature it is possible to observe the populations of the $\Ket{+1/2}$ or $\Ket{-1/2}$ states through their interaction with the cavity modes. The results can be represented in terms of the effective susceptibility of the medium that is proportional to the number of impurities which are able to interact with the resonance of a certain polarisation. 




The DC susceptibility can be found from the experimental data of the avoided crossing between the resonator WGMs and Fe$^{3+}$ electron spin resonance (ESR) using Eqn.~(\ref{B002SF}):
\begin{equation}
	\label{B004SFa}
	\chi_\pm = \frac{g_\pm^2}{\omega_0^2\xi},
\end{equation}
where $\omega_0$ is either $\omega_R$ or $\omega_L$, $\xi$ is the filling factor for the modes calculated for the magnetic field perpendicular to the external DC field, i.e. the sum of the radial and azimuthal fields. In order to find the coupling coefficients $g_\pm$ one needs to consider the simplified model for the frequencies in the vicinity of the given transition \hbox{($+$ or $-$):}
\begin{multline}
	\label{B004SF}
\displaystyle  H = \omega_R\hbar a_R^\dagger a_R +  \omega_L\hbar a_L^\dagger a_L + g_{RL}\hbar\Big(a_R^\dagger a_L+ a_R a_L^\dagger\Big)\\
\displaystyle+\hbar\frac{\omega_+}{2}\sum_{i=1}^{N_+}\sigma^z_{+i} 
\displaystyle +\hbar \widetilde{g}_+\sum_{i=1}^{N_+}\big(\sigma_{+i}^+a_L+\sigma_{+i}^-a_L^\dagger\big),
\end{multline}
where one of the sub-ensembles is neglected (in this case the `minus' transition is neglected), since the corresponding type of interaction is prohibited by the spin angular momentum and energy conservation laws. Thus, in the given case only LHC photons are coupled to the spin bath, although RHC photons are still coupled to the LHC resonance indirectly due to the existence of backscatterers. Eqn.~(\ref{B004SF}) is effectively a three-oscillator model since an unsaturated ensemble of TLSs could be considered as another bosonic degree of freedom\cite{ritsch}. Note that whenever $g_\pm\gg g_{RL}$, the second and the third terms in the Hamiltonian (\ref{B004SF}) can be neglected, and only two HOs (one photon and one spin) need to be considered. Such a two-HO model is effectively a Jaynes-Cummings system. However, as demonstrated below this model is not valid for all temperatures.

In the three-oscillator model, the angular frequencies of the RHC and LHC polarised resonances ($\omega_R$ and $\omega_L$), as well as the effective ESR angular frequency tuned by the magnetic field ($\omega_+$) can be identified from the experimental data for the `plus' mode interaction. The coupling parameters $g_{RL}$ and $g_+=\widetilde{g}_+\sqrt{N_+}$ can be fitted in such a way that model is in good agreement with the measured avoided crossing. As described above, the latter parameter allows the calculation of the DC susceptibility $\chi_+$ given by Eqn.~(\ref{B004SFa}), which represents the effective response of the medium to the magnetic component of the wave. This parameter demonstrates a clear temperature dependence, since the change of the crystal temperature leads to change of size of each sub-ensemble $N_{+}$, in particular modifying the upper summation limit in the forth term of the Hamiltonian (\ref{B004SF}) as described by Eqn.~(\ref{B004aSF}).


\begin{figure}[ht!]
	\centering
			\includegraphics[width=0.50\textwidth]{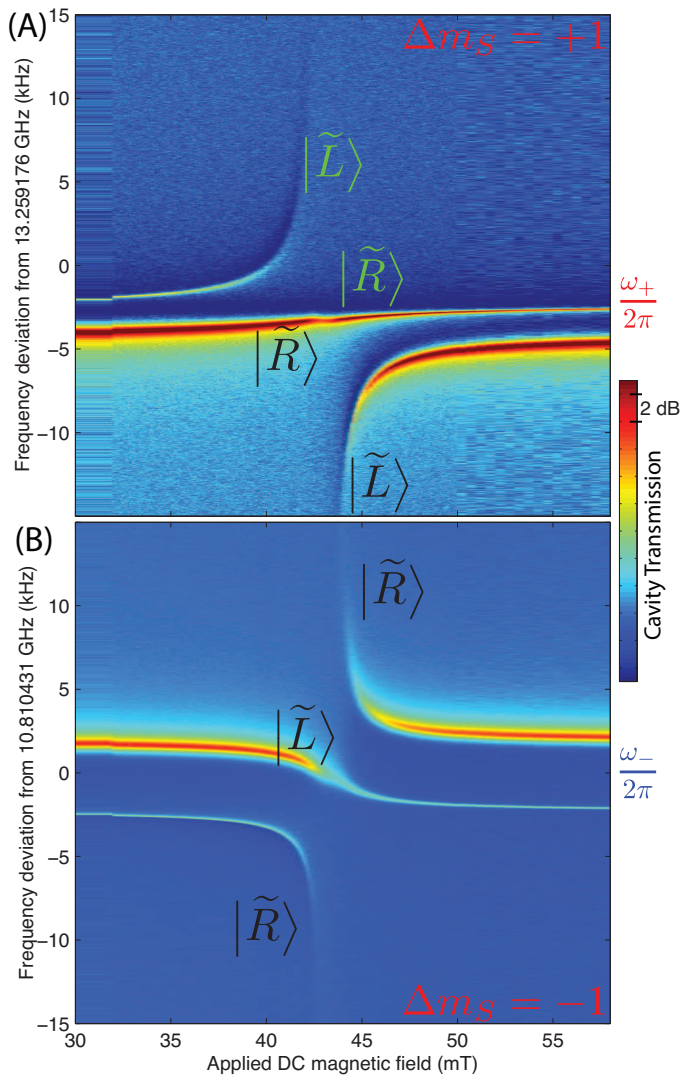}
	\caption{Weak interaction of the ensemble of Fe$^{3+}$ ions with two WGM doublets in the transitional case $g_{\pm}\rightarrow g_{RL}$ ($T\sim 5$ K). (A) shows the case of a spin-increasing ion transition $\omega_+$,  (B) shows the case of a spin-decreasing transition.}
	\label{plus2}
\end{figure}

\begin{figure}[ht!]
	\centering
			\includegraphics[width=0.50\textwidth]{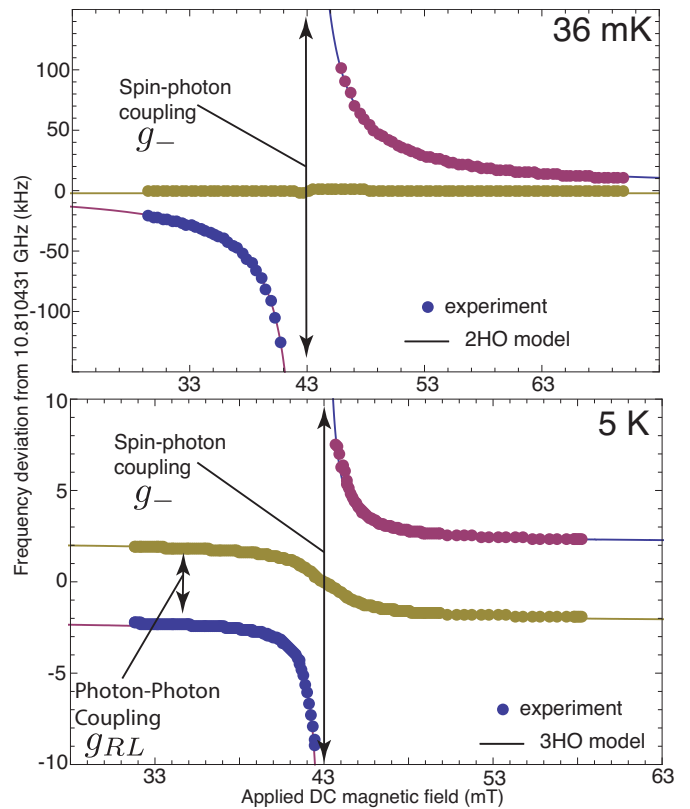}
	\caption{Comparison of the spin-dominated regime ($T=36$~mK, $g_{\pm}\gg g_{RL}$) and the spin regime with significant backscatter ($5$ K, $g_{\pm}\rightarrow g_{RL}$). The former interaction can be modelled by two oscillator model (one WGM and one spin ensemble). The latter requires a three oscillator model (two for WGMs, and one for the spin ensemble). }
	\label{model}
\end{figure}

\begin{figure}[ht!]
	\centering
			\includegraphics[width=0.50\textwidth]{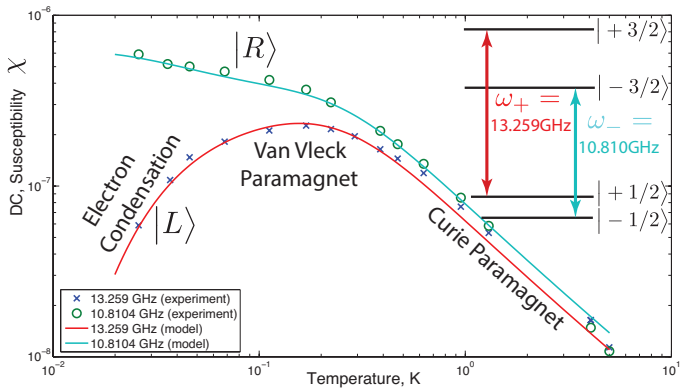}
	\caption{Measured and predicted susceptibilities of two WG modes $f_A$ and $f_B$ coupled to two transitions $\omega_+$ and $\omega_-$.}
	\label{susept}
\end{figure}

The identified temperature dependence of the DC susceptibility is shown in Fig.~\ref{susept}.
Decreasing the crystal temperature sees an increase in population of both the $\Ket{+1/2}$ and $\Ket{-1/2}$ states due to the condensation of ions mainly from the $\Ket{\pm3/2}$ states. This is the well-known Curie Law, which we observed at temperatures between $0.2$ and $5$K with a dependence of $\chi\sim T^{-1-|\varepsilon|}$. The existence of the $\Ket{+5/2}$ and $\Ket{-5/2}$ levels leads to a small deviation $|\varepsilon|$ from this law for temperatures above $4$K. At such temperatures, the spin-photon coupling is reduced to the level where the HOs coupling between the WGM doublets $g_{RL}$ becomes important. In this case, when $g_{-}$ or $g_{+} \rightarrow g_{RL}$ (depending on the sign of the spin change), the interaction picture (Fig.~\ref{plus2}) becomes significantly different from that demonstrated in Fig.~\ref{plus}. In particular, the reduced two oscillator model becomes invalid. The system requires the full three oscillator description (Fig.~\ref{geom} and Hamiltonian~(\ref{B002SF})). In the extreme case when either $g_{\pm} < g_{RL}$, it becomes impossible to distinguish between polarisations due to the weak coupling of the spin to the photons. In this case, the doublet WGMs remain as standing waves in a similar way to the Fabry-P\'{e}rot cavity mentioned in the Introduction section. 

Comparing the strong spin-photon interaction case at millikelvin temperatures, when one of $g_{-}$ or $g_+  \gg g_{RL}$ as shown in Fig.~\ref{plus}, with the weaker spin-photon interaction case at 5K, when one of $g_{-}$ or $g_+  \rightarrow g_{RL}$, with the other equal to zero (Fig.~\ref{plus2}), it can be noted that whereas in the former case one of the resonances of the doublet is only slightly perturbed, in the latter case both resonances are significantly displaced (although one more than the other). Thus, in the latter case it is harder to identify each of the HOs with pure $\Ket{R}$ or $\Ket{L}$ polarisations, since the backscattering mechanism mixes the states of photon spin angular momentum to some degree. In other words, in this situation the cavity photon behaviour is influenced by the backscattering mechanism as well as the spin-photon coupling. Thus, the former case can be described as a spin-dominated regime while the latter demonstrates a transitional case to a backscatter domination regime . The two regimes are compared in Fig.~\ref{model} with the experimental data compared to two- and three-HO models respectively for the two regimes described above.

For temperatures equal to and below $0.2$K, the increase of spin-photon coupling saturates as the temperature decreases, due to the fact that the upper levels become empty and thus a further increase of susceptibility becomes impossible. Thus, this paramagnetic phase has no increase of susceptibility with decreasing temperature and is known as a Van Vleck paramagnet\cite{vleck}. For even lower temperatures, ions at the $\Ket{+1/2}$ level condense to the lowest energy state $\Ket{-1/2}$ leading to a further increase of susceptibility for RHC polarised photons, and a decrease of susceptibility for LHC polarised photons. Theoretically, the susceptibility is calculated to be proportional to the number of electrons in a state leading to a certain type of interaction:
\begin{equation}
	\label{FD02SF}
	\left. \begin{array}{ll}
\displaystyle  \chi_{\pm} \sim  \frac{1}{Z}\Big(e^{\beta E_{\Ket{\pm1/2}}}-e^{-\beta E_{\Ket{\pm1/2}}}\Big)
\end{array} \right.
\end{equation}
where  the partition function is found taking into account all accessible spin states:  $Z = \exp\Big(-\beta E_{\Ket{q}}\Big)$,
where $q$ runs over $\pm1/2$, $\pm3/2$ and $\pm5/2$. Fig.~\ref{susept} shows that the susceptibility indeed varies as predicted by the ion thermal distribution model. 

\section{Conclusion}

The experiments presented here reveal the significant role of photon polarisation states\cite{raymer}  on wave-matter interactions in WGM dielectric resonators\cite{paperA}. In particular, the situation requires consideration of the spin angular momentum conservation law in addition to the usual energy conservation. These additional features on top of the usual Jaynes-Cummings type interactions in a Fabry-P\'{e}rot type cavity arise due to non-degeneracy of right and left hand polarised photons. The latter fact is the direct consequence of {both reflection and time-reversal symmetry breakings} in an actual cylindrical-type cavity. The effects discussed are only observable at the spin dominated regime where spin-photon coupling is much stronger than backscattering coupling between the resonances of a doublet.

\section*{Acknowledgements}
This work was supported by Australian Research Council grants CE110001013 and FL0992016.

\section*{References}

%


\end{document}